# Can the neutrino speed anomaly be defended?

J. Knobloch/CERN

**Abstract**

The OPERA collaboration reported [1] a measurement of the neutrino velocity exceeding the speed of light by 0.025‰. For the 730 km distance from CERN in Geneva to the OPERA experiment an early arrival of the neutrinos of 60.7 ns is measured with an accuracy of ±6.9 ns (stat.) and ±7.4 ns (sys.). A basic assumption in the analysis is that the proton time structure represents exactly the time structure of the neutrino flux. In this manuscript, we challenge this assumption. We identify two main origins of systematic effects: a group delay due to low pass filters acting on the particular shape of the proton time distribution and a movement of the proton beam at the target during the leading and trailing slopes of the spill.

**Introduction**

In a speed measurement there are two major elements: distance and time. In the OPERA analysis there is, however, a third element that did not give rise to detailed consideration in [1]: The measurement of the time structure or Particle Density Function (PDF) of the neutrinos emanating from the CERN CNGS (CERN Neutrinos to Gran Sasso) system. The proton extraction lasts for 10.5 μs and there are 16111 neutrino events in OPERA used in the analysis. The statistical accuracy (for a rectangular PDF) would be
$\Delta = 10.5\ \mu s/\sqrt{12}*\sqrt{16111} = 24$ ns. The claimed anomaly of 60.7 ns is, however, measured more precisely with an accuracy of ±6.9 ns (stat.) and ±7.4 ns (sys.). Therefore the leading and trailing edges of the neutrino time distribution play an important role in the analysis. OPERA assumes that the proton PDF is measured correctly and that it represents exactly the neutrino PDF. In the following, we argue that both assumptions can be questioned and that systematic effects of the order of the observed anomaly have been neglected.

**The neutrino beam**

In the CERN Super Proton Synchrotron (SPS), 400 GeV protons complete one round trip in 23 μs. The ring is filled with two proton batches of 10.5 μs separated by two gaps of 1 μs. The batches are extracted spaced by 50 ms. The extraction of each batch is initiated by a kicker magnet powered up with a 1.1 μs rise time during the 1 μs gap. Once the kicker magnet strength reaches 80% of the maximum, the beam trajectory is inside the gap of a magnetic septum at the beginning of a beam line leading to a 4 mm (and 5 mm) diameter carbon target where charged mesons are produced that are subsequently charge selected and focused by a magnetic horn/reflector system. In a 1 km decay tunnel the mesons decay into muons and muon neutrinos targeted at the OPERA experiment 730 km away.
Just after the septum, 743 m upstream of the target, a beam current transformer (BCT) measures the proton flux in a coil coaxial with the beam. The signal of the BCT is amplified and transported by a 140 m long cable to a precision waveform analyser (WFA).

**Correct measurement of the proton PDF**

In a thesis [2], additional details of the analysis are given. For certain run periods, the digitizer did not perform correctly by either saturating the signal or by inducing oscillations. These periods have been removed from the analysis. Assuming that all such periods could be traced, there remains, however, an oscillating 30 and 60 ns structure in the measured waveforms, most pronounced during the last quarter of the extractions and in particular over the falling edge of the proton spill, see Fig. 8.4 of [2].
Such oscillations are still visible after summing 16111 individual measurements.
In the analysis described in [2], the oscillations are eliminated by a low-band software filter of 8 MHz. A low-pass filter not only attenuates the noise but also inflicts a frequency dependent group delay. The filter algorithm used was not specified; therefore we have evaluated the group delay curve for several low-pass filters using FilterDesignLab II-R [3]. As example Fig. 1 shows the group delay curve for an 8 MHz Butterworth low-pass filter.

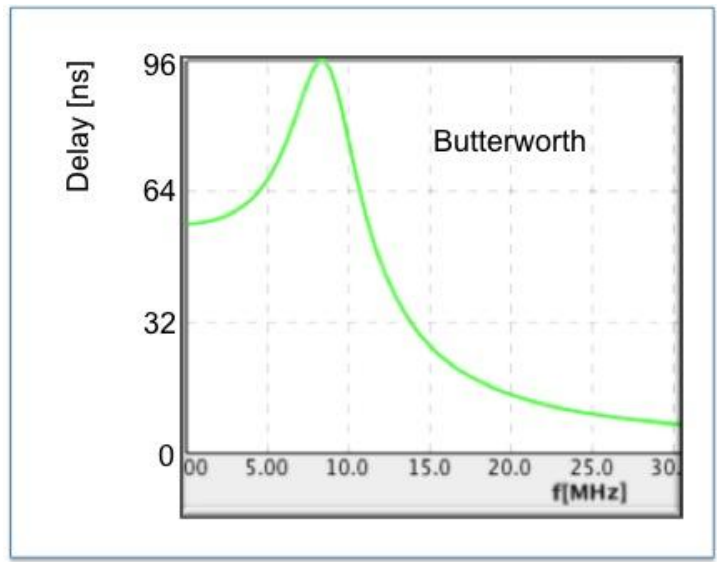

Figure 1: Group delay for 8 MHz low-pass filter

The proton PDF has a leading edge rise time of about 800 ns and a trailing edge fall time of about 400 ns (see Fig. 12 of [1]) and a more or less flat top of 9300 ns in between. The characteristic frequency of such a pulse shape is given by the rise and fall time $\tau \approx 600$ ns and amounts to $f = 1/(2\pi\tau) \approx 0.3$ MHz. In the group delay curve in Fig. 1 this frequency is at the left side of the distribution and leads to a group delay of about 60 ns. Note that a nanosecond short pulse would be located at the right side of the distribution and leads to a comparatively small group delay. Different kinds of 8 MHz low-pass filters (Chebychev-1, Chebychev-2, elliptic) have been evaluated yielding group delays between 30 and 60 ns.
A similar, though smaller, effect can impede the measurement of the time delay from the BCT to the waveform digitizer. The system BCT-amplifier-transmission cable constitutes a low-pass filter with a cut-off frequency of about 80 MHz [4][1]. The delay is measured using either a 1 PPS pulse from a Cs4000 oscillator or, with better precision, short (nanosecond) proton bunches. Both signals have short (< 5 ns) rise times. For such signals the high-frequency part of the group delay curve is relevant. The proton waveform is located at lower frequencies of 0.3 MHz. The net delay due to this effect is of the order of 10 ns.

---

[1] The low pass filter effect can also be seen in Fig. 4 of [1] where the 200 MHz SPS RF structure is attenuated by 70% corresponding to a 10.5 dB attenuation.

**Broadening of the PDF**

The proton PDF is a sum of the individual BCT measurements that are coincident with neutrino events in OPERA. In the summing process, the time alignment of the individual distributions is based on the trigger signal of the kicker magnet MKE. The timing of this trigger (kick delay) is occasionally optimized (in steps of 100 ns) in order to minimize beam loss, in particular in the septum magnet. After such optimization or after a machine development period this delay may not come back to the previous value. If this would happen during the yearly data taking, some fraction of the proton distributions would be shifted by e.g. 100 ns. Such effect would lead to a broadening of the summed PDF.
It can therefore not be excluded that the widths of the used proton PDF and the neutrino event distribution are different. An indication of such broadening could be that the single waveform in Fig. 4 of [1] appears to have steeper edges than the sum shown in Fig. 12 of [1].
One may argue that this broadening would not change the mean of the distribution and therefore not impact the result. As mentioned above, however, the leading and trailing slopes differ by about a factor 2. As the steeper slope will have a larger impact on the fit result, this will lead to a shift in the final measurement. Visually, Fig. 12 of [1] does not allow excluding a broadening of the PDF by the order of 40 ns.

**Differences between proton and neutrino PDF**

The proton PDF is measured with a BCT 730 m upstream of the target. The neutrino PDF is proportional to the proton PDF only if all the protons measured in the BCT actually hit the target and if the position of the beam at the target does not move during the spill.
The gap of 1 µs between the proton batches in the SPS is not completely void of protons and therefore during the final 20% of the kicker magnet ramp, where the beam is within the acceptance of the neutrino beam line, protons are extracted and counted in the BCT (see [5] and [6]). The 20% kicker variation translates to a 10 mm beam movement at the septum. This can lead to a beam movement at the target of about 2 mm.
The beam position stability at the target is claimed to be 50 µm r.m.s. in the vertical plane and 100 µm r.m.s. in the horizontal plane [7]. Due to long integration time of the beam monitors, however, this measurement reflects the beam properties during the 9.3 µs long spill plateau and cannot account for deviations at the leading and trailing edges. The beam spot is 0.5 mm rms.
The effect on the neutrino flux of a beam displacement at the target has been studied [7] and is shown in Figure 2. From this figure it is clear that a 1.5 to 2 mm displacement will lead to a significant reduction of the neutrino yield during the leading and trailing edges of the PDF and therefore distort their shapes.

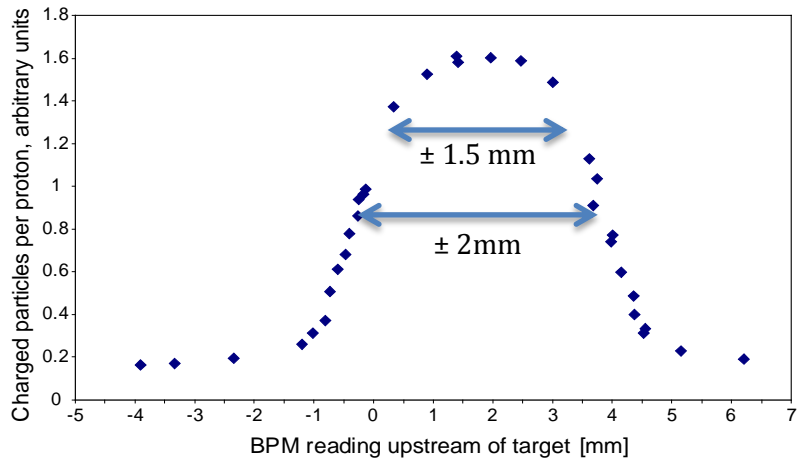

Figure 2: Horizontal proton position scan: number of charged particles in arbitrary units produced for different proton beam positions onto the target, as read by the BPM upstream of the target. (Data presented in [7]).

**Conclusion and outlook**

We have seen that
- The OPERA result critically depends on the edges of the PDF.
- The measurement of the proton PDF is subject to bandwidth limitation or low-pass filtering leading to group velocity delay.
- The leading slope of the individual proton time distribution is steeper than that of the final PDF.
- During the final 20% of the ejection kicker ramp up there are already protons ejected.
- The variation of kicker strength leads to 10 mm movement of the beam in the CNGS beam line
- The movement may displace the beam at the target by 1-2 mm.
- These effects act predominantly on the leading and trailing edges of the PDF

We conclude that a difference between the proton and neutrino PDF was not sufficiently considered in evaluating the systematic uncertainties summarized in Table 2 of [1]. The effects discussed in this present manuscript amount to a significant fraction of the observed anomaly. Therefore a conclusive deviation of the neutrino velocity from the speed of light has not yet been demonstrated in [1].
A new measurement of the neutrino speed is being conducted using 2 ns long bunches spaced by 500 ns from the CNGS beam [8]. In such a measurement the effects discussed here would no longer apply.

**Acknowledgements**

The author thanks F. Dydak for useful comments and encouragement. D. Belohrad has provided details of the BCT and I. Efthymiopoulos has discussed details of the neutrino beam. In the preparation of this manuscript, slides of presentations by J. Wenninger and E. Gschwendtner were most useful. Not least, the hospitality of the Pontifical Academy of Sciences and the support of A. Zichichi have made this work possible.